\documentclass[twocolumn,showpacs,preprintnumbers,amsmath,amssymb]{revtex4}

\usepackage{graphicx}
\usepackage{dcolumn}
\usepackage{bm}
\begin{document}
\title{Quantization of the metric diagonal spacetime with Gaussian normal coordinates}

\author{Shintaro Sawayama}
 \email{shintaro@th.phys.titech.ac.jp}
\affiliation{Department of Physics, Tokyo Institute of Technology, Oh-Okayama 1-12-1, Meguro-ku, Tokyo 152-8550, Japan}
\begin{abstract}
In the analysis of the Wheeler-DeWitt equation, we have simplified the Hamiltonian constraint of the Wheeler-DeWitt equation 
using the coordinate transformation.
The coordinate is choose such that metric becomes diagonal and as Gaussian normal coordinate.
Or we treat small universe so that the metric become diagonal and universe is covered by Gaussian normal coordinates.
We have solved the Wheeler-DeWitt equation of such universes.
Such that universe contains Biancki I type universe or the black hole universe.
\end{abstract}

\pacs{04.60.-m, 04.60.Ds}
\maketitle

\section{Introduction}
In the theory of the quantum gravity there are many approach such that string \cite{AL} or loop \cite{Rov}\cite{As}\cite{Thi} or mini-super approach \cite{Hart}.
However, the theory of the quantum gravity has not completed yet.
Mainly the quantum gravity based on the ADM decomposition \cite{ADM}.
Form the ADM decomposition; we obtain constraint equations i.e. Hamiltonian and diffeomorphism constraint equations.
To solve these constraint equations is the orthodox way of the canonical quantum gravity.
The Hamiltonian constraint is the generator of the time translation and the diffeomorphism constraint are the generator of space translations \cite{Di2}. 
The theory of quantum gravity contains many unsolved problems which contains problem of the time and problem of the norm.
However, most important problem is the difficulty of the constraint equations, i.e. Wheeler-DeWitt equation \cite{De}.
There remain problem of the norm or problem of diffeomorphism and problem of the time.
The main difficulty is in the Wheeler-DeWitt equation.
By the our resent study \cite{Sa} we can mention for problem of the time.
The theory of the quantum gravity start when the Wheeler-DeWitt equation is solved.

We think that the theory stats with solving the inhomogeneous spacetime.
So we treat the quantization of the inhomogeneous spacetime.
We would like to know at least one inhomogeneous quantum state.
As a result we have solved one Wheeler-DeWitt equation.
Although our model is metric diagonal case and the Gaussian normal coordinate condition,
the application is large enough to contain the enlargement of the Biancki I type model and the black hole model
.
In this paper we treat the small universe whose metric can be chosen diagonal by coordinate transformation and whose coordinate is chosen as the Gaussian normal.
Then we can solve the Hamiltonian constraint.
Secondly we recover the off-diagonal components and solve the diffeomorphism constraint.
However, we do not cover this second step.

\section{Simplification of the Hamiltonian constraint}
Using the fact that the metric become diagonal by the local coordinate transformation,
we start from decomposition of the Einstein Hilbert action as
\begin{eqnarray}
S=\int RdM=\sum_i \int R_i[g_{\mu\mu}]dSdt.
\end{eqnarray}
Here $S$ is the hyper-surface with constant time.
And $S$ is defined such that metric become diagonal by the local coordinate transformation.
And from $S$ the causal diamond covers all most universe.
Because, our method is different from the usual Wheeler-DeWitt equation formalism,
our obtained Hamiltonian constraint is different type of the Wheeler-DeWitt equation.
If we decompose this action as 3+1, then we can obtain
\begin{eqnarray}
{\cal L}=\dot{q}_{ii}P^{ii}+NH-2\sqrt{q}D^iN_{,i}.
\end{eqnarray}
Here $N$ is the lapse functional and $H$ is the Hamiltonian constraint such that
\begin{eqnarray}
H=\frac{1}{2}q_{ii}q_{jj}P^{ii}P^{jj}+{\cal R}.
\end{eqnarray}
Here ${\cal R}$ is the three dimensional Ricci scalar and $P^{ii}$ is the momentum whose commutation relation with $q_{ii}$ is not $i$, but $i\sqrt{q}$.
In this formulation there are not appear $q_{ij}$ and $P^{ij}$ and sift vectors and momentum constraint.
So we can ignore the constraint as $[P^{ij},H]$, because we start with metric diagonal setting.
If we write the Hamiltonian constraint in the operator representation, we obtain
\begin{eqnarray}
H=\sum_{ij}\frac{1}{2}\frac{\delta^2}{\delta \phi_i\delta \phi_j}+
{\cal R}[q_{11},q_{22},q_{33}]=0 \nonumber \\
=\sum_{ij}\frac{1}{2}\frac{\delta^2}{\delta \phi_i\delta \phi_j}+
\sum_{i\not= j}(\hat{\phi}_{i,jj}+\hat{\phi}_{j,i}\hat{\phi}_{i,i})e^{\hat{\phi}_i}=0.
\end{eqnarray}
Because the final term comes from $\Gamma\Gamma$ term, if we used Gaussian normal coordinate,
the Hamiltonian constraint becomes
\begin{eqnarray}
H\to \sum_{ij}\frac{1}{2}\frac{\delta^2}{\delta \phi_i\delta \phi_j}+
\sum_{i\not= j}\hat{\phi}_{i,jj}e^{\hat{\phi}_i}=0.
\end{eqnarray}
Here $\Gamma$ means the Christoffel symbol.
We start with this simplified model.
We comment on the final step.
The assumption such that $\Gamma |\Psi\rangle=0$ is incorrect.
However, we start with Gaussian normal coordinate,
the $\Gamma \Gamma$ terms drooped out in the derivation of ${\cal R}$.

\section{Solving the Wheeler-DeWitt equation}
Solving the above Hamiltonian constraint, we use static restriction at the first time as a ansatz and 
next time we calculate the solution without assumption.
If we assume ansatz which we call static restriction such as
\begin{eqnarray}
\sum_{i\not= j}\frac{\delta^2}{\delta\phi_i\delta\phi_j}=0,
\end{eqnarray}
Then this constraint and the Hamiltonian constraint commute and we can quantize simultaneously.
Usually the static restriction and the Hamiltonian constraint does not commute.
However, if we use a Gaussian normal coordinate, static restriction commute with the Hamiltonian constraint.
Using the static restriction we can simplify the Hamiltonian constraint as
\begin{eqnarray}
\sum_i\frac{\delta^2}{\delta \phi_i^2}+2\phi_{i,jj} e^{\phi_i}=0.
\end{eqnarray}
If we assume the state is parameter separated, the Hamiltonian constraint equation is reduced to
\begin{eqnarray}
\frac{\delta^2}{\delta a_i^2}+8\partial^j\partial_j\ln \hat{a_i}=0 \ \ \ {\rm for \ \ some} \ \ i
\end{eqnarray}
Here, $a_i=g_{ii}^{1/2}$ and hat means operator.
Using the approximation as
\begin{eqnarray}
\partial^j\partial_j\ln \hat{a_i}\frac{\delta}{\delta a_i}-\frac{\delta}{\delta a_i}\partial^j\partial_j\ln \hat{a_i}
\end{eqnarray}
is small, we can solve the Hamiltonian constraint as
\begin{eqnarray}
\exp (2\sqrt{2}i\int (\partial^j\partial_j\ln a_i)^{1/2}\delta a_i) \ \ \ {\rm for} \ \ \  i\not= j.
\end{eqnarray}
Then the state becomes
\begin{eqnarray}
\exp (2\sqrt{2}i\sum_i\int (\partial^j\partial_j\ln a_i)^{1/2}\delta a_i) \ \ \ {\rm for} \ \ \  i\not= j.
\end{eqnarray}
Because of the static restriction there is a gauge in the coordinate
\begin{eqnarray}
\sum_{i,k}\sum_i(\partial^j\partial_j\ln a_i)^{1/2}\sum_k(\partial^j\partial_j\ln a_k)^{1/2}=0.
\end{eqnarray}
Because of this special gauge the above state may be empty.
And because of this gauge, we obtain one of the $g_{i,jj}$ by the other two $g_{i,jj}$.
However, if we assume the solution of Eq.(5) is form of the 
\begin{eqnarray}
|\Psi\rangle =\exp (2\sqrt{2}i\sum_{i}\int (\sum_j\partial^j\partial_j\ln a_i)^{1/2}\delta a_i)f[\phi] ,
\end{eqnarray}
we can solve Eq.(5).
If we act the above state to Eq.(5), we obtain
\begin{eqnarray}
\sum_{i,j}\frac{\delta}{\delta \phi_i}\bigg(\frac{\delta f}{\delta \phi_j}\bigg)+\sum_k 2\phi_{i,kk}\frac{\delta f}{\delta\phi_j}
\end{eqnarray}
If we write
\begin{eqnarray}
\nabla =\sum_i\frac{\delta}{\delta \phi_i}
\end{eqnarray}
and 
\begin{eqnarray}
\nabla f=g,
\end{eqnarray}
we can write the Eq.(14) as
\begin{eqnarray}
\nabla g+2\sum_j\phi_{i,jj}g=0.
\end{eqnarray}
This equation can be solved with similar technique to derive Eq.(11) and solution is of the form
\begin{eqnarray}
g=\exp \bigg(2\sum_{i}\int \sum_j\phi_{i,jj}\delta \phi_i \bigg)
\end{eqnarray}
And we can write $f[\phi]$ symbolically as
\begin{eqnarray}
f[\phi]=\nabla^{-1}\exp \bigg(2\sum_i\int \sum_j\phi_{i,jj}\delta \phi_i \bigg) ,
\end{eqnarray}
and the state is written as
\begin{eqnarray}
|\Psi^4(q)\rangle =\exp (2\sqrt{2}i\sum_i\int (\sum_j\partial^j\partial_j\ln a_i)^{1/2}\delta a_i) \nonumber \\
\times \nabla^{-1}\exp \bigg(2\sum_i\int \sum_j\phi_{i,jj}\delta \phi_i \bigg)
\end{eqnarray}
The above solution is the main result of our work.
We know the state is second integrated by metrics from the state.
It is same to treat following metric such as
\begin{eqnarray}
ds^2=\begin{pmatrix}
-N^2 & 0 & 0 & 0 \\
0 & q_1(t,x_1,x_2,x_3) & 0 & 0 \\
0 & 0 & q_2(t,x_1,x_2,x_3) & 0 \\
0 & 0 & 0 & q_3(t,x_1,x_2,x_3) ,
\end{pmatrix}
\end{eqnarray}
with the restriction as
\begin{eqnarray}
\Gamma^{(3)i}_{ij}=\frac{1}{2}q^{ii}q_{ii,j}=0,
\end{eqnarray}
where $\Gamma^{(3)}$ is the three dimensional Christoffel symbol.
In terms of $\phi$, this constraint can be written as
\begin{eqnarray}
\sum_j \phi_{i,j}=0.
\end{eqnarray}
If we assume $\phi=\phi_1=\phi_2=\phi_3$, the state become simple as
\begin{eqnarray}
|\Psi^4(q)\rangle =\exp (6\sqrt{2}i\int (\sum_j\partial^j\partial_j\ln a)^{1/2}\delta a) \nonumber \\
\times \int \exp \bigg(6\int \sum_j\phi_{,jj}\delta \phi \bigg) \delta \phi .
\end{eqnarray}
\section{Diffeomorphism constraint}
From the Eq.(20) we recover the off-diagonal components of the metrics by usual sense such as
\begin{eqnarray}
\phi_i\to \sum_{i'j'}\phi_{i'}\frac{\partial x_i}{\partial x_{i'}}\frac{\partial x_i}{\partial x_{j'}}
\end{eqnarray}
\begin{eqnarray}
\phi_{i,kk}\to \sum_{i'j'}\phi_{i',k'k'}\frac{\partial x_{k'}}{\partial x_i}\frac{\partial x_{k'}}{\partial x_i}
\frac{\partial x_i}{\partial x_{i'}}\frac{\partial x_i}{\partial x_{j'}} \nonumber \\
=\sum_{i'j'}\phi_{i',k'k'}
\frac{\partial x_{k'}}{\partial x_{i'}}\frac{\partial x_{k'}}{\partial x_{j'}} \nonumber \\
=\phi_{i',k'k'}
\end{eqnarray}
Then the Eq.(20) becomes as
\begin{eqnarray}
|\Psi^4(q)\rangle =\exp (2\sqrt{2}i\sum_{i'}\int (\sum_{j'}\partial^{j'}\partial_{j'}\ln a_{i'})^{1/2}\delta a_{i'}) \nonumber \\
\times \nabla^{-1}\exp \bigg(2\sum_{i'}\int \sum_{j'}\phi_{i',j'j'}\delta \phi_{i'} \bigg)
\end{eqnarray}
Here, we used the $\nabla^{-1}$ does not change because of integration.
From the above equation, we know the solution is diffeomorphism invariant.
Although there do not appear the diffeomorphism constraint at first in this model,
the state satisfy the diffeomorphism constraint automatically.
\section{Conclusion and discussions}
We have solved Wheeler-DeWitt equation with restricting the coordinate so that the metric become diagonal and as Gaussian normal coordinate under special gauge.
The know that the state has a form of second integrated as expected.
And the obtained state automatically satisfy the diffeomorphism constraint.
We know the states are only depend on $\phi_{i,jj}$.
Although our obtained result is complicated because of the mixed integration, 
we can simplify the state so that it can be written only simple integration.
In the solving step we assume the state is parameter separated.
So our obtained result is not the overall state.

Although we stopped the study at the solving step, the obtained state can be applied to enlarged Biancki I universe or the black hole universes.
Although the $\nabla^{-1}$ is some complicated integration,
we can make it simple integration by becoming the metric as $q_{11}=q_{22}=q_{33}$.

Because we found at least one inhomogeneous state,
we can study the problem of the norm.
Or we can study the fluctuation of the CMB.
And we may comment on the singularity problem.

Our final goal is the quantization of the black holes.
However, there are many problems to solve it.

\end{document}